\documentclass[preprint2]{aastex}

\slugcomment{to appear in ApJ}

\shorttitle{O isotopic ratios cool RCB stars}
\shortauthors{Garc\'{\i}a-Hern\'andez et al.}

\begin{document}

\title{Oxygen isotopic ratios in cool R Coronae Borealis stars}

\author{D. An\'\i bal Garc\'\i a-Hern\'andez\altaffilmark{1}, David L.
Lambert\altaffilmark{2}, N. Kameswara Rao\altaffilmark{3}, Ken H. Hinkle\altaffilmark{4}, Kjell Eriksson\altaffilmark{5}}
\altaffiltext{1}{Instituto de Astrof\'{\i}sica de Canarias, C/ Via L\'actea
s/n, 38200 La Laguna, Spain; agarcia@iac.es}
\altaffiltext{2}{W. J. McDonald Observatory. The University of Texas at
Austin. 1 University Station, C1400. Austin, TX 78712$-$0259, USA; dll@astro.as.utexas.edu}
\altaffiltext{3}{Indian Institute of Astrophysics, Bangalore 560034, India; nkrao@iiap.res.in}
\altaffiltext{4}{National Optical Astronomy Observatory (NOAO), Tucson, AZ85726,
USA; hinkle@noao.edu}
\altaffiltext{5}{Department of Physics and Astronomy, Uppsala University,
Box 515, 75120 Uppsala, Sweden; Kjell.Eriksson@astro.uu.se}

\begin{abstract}

We investigate the relationship between R Coronae Borealis (RCB) stars and
hydrogen-deficient carbon (HdC) stars by measuring precise $^{16}$O/$^{18}$O
ratios for five cool RCB stars. The $^{16}$O/$^{18}$O ratios are derived by
spectrum synthesis from high-resolution (R$\sim$50,000) K-band spectra. Lower
limits to the $^{16}$O/$^{17}$O and $^{14}$N/$^{15}$N ratios as well as Na and S
abundances (when possible) are also given. RCB stars in our sample generally
display less $^{18}$O than HdC stars - the derived $^{16}$O/$^{18}$O ratios
range from 3 to 20. The only exception is the RCB star WX CrA, which seems to be
a HdC-like star with $^{16}$O/$^{18}$O=0.3. Our result of a higher
$^{16}$O/$^{18}$O ratio for the RCB stars must be accounted for by 
a theory of the formation and evolution of HdC and RCB stars.
We speculate that a late dredge-up of products of He-burning, principally
$^{12}$C and $^{16}$O, may convert a $^{18}$O-rich HdC star into a $^{18}$O-poor
RCB star as the H-deficient star begins its final evolution from a cool
supergiant to the top of the white dwarf cooling track.

\end{abstract}

\keywords{stars: abundances --- stars: atmospheres  --- stars: chemically
peculiar --- stars: white dwarfs --- infrared: stars}

\section{Introduction}

The R Coronae Borealis (RCB) stars and their likely cousins the
hydrogen-deficient carbon (HdC) stars and the extreme helium (EHe) stars have
long posed a puzzle as to their origins in terms of stellar evolution. Two
leading scenarios have survived decades of theoretical and observational
scrutiny. In one, the H-deficient supergiant is formed from the merger of a He
white dwarf (WD) with a C-O white dwarf (Webbink 1984; Iben \& Tutukov 1984;
Saio \& Jeffery 2002). This path is widely referred to as the
double-degenerate (DD) scenario. In the other, these H-deficient stars result
from a final, post-AGB helium shell flash in the central star of a planetary
nebula. The final flash may transform the star on the WD cooling track into a
H-deficient supergiant; this the so-called ``born-again" scenario is discussed
by Herwig (2001) and Bl\"{o}cker (2001) and often labeled the FF scenario. Open
questions remaining include: Is the DD or the FF scenario the dominant
mechanism? If both are operative, how does one distinguish a product of the DD
from one of the FF scenario?

Elemental and isotopic abundances of C, N, and O are potentially powerful agents
for testing the different evolutionary scenarios proposed. Previous studies of
the CNO abundances support production of the RCBs and the EHes by the DD
scenario (Pandey et al. 2006). A dramatic advance was made by Clayton et al.'s
(2005, 2007) discovery from medium-resolution infrared spectra of the CO 2.3
$\mu$m bands that $^{18}$O was very abundant for some cool RCBs and HdCs.
Exploratory calculations led Clayton et al. (2007) to propose that $^{18}$O was
synthesised from $^{14}$N during the merger in the DD scenario. Extraordinary
conditions are required for the FF scenario to lead to abundant $^{18}$O. 

More recently, we analyzed high-resolution (R=50,000) spectra of a few narrow
windows in the K band of the five known HdC stars and a few RCB stars (Garc\'\i
a-Hern\'andez et al. 2009). In these spectra, the CO spectrum is resolved and
application of spectrum synthesis enables a more precise estimate of the
$^{16}$O/$^{18}$O ratio to be obtained. Our recent analysis generally confirmed
reports by Clayton and colleagues from R=5,900 (or lower resolution) spectra. We
confirm that the $^{16}$O/$^{18}$O ratio is less than unity for those HdC stars
(3 of the known 5) exhibiting CO lines in their spectra.  However,
$^{16}$O/$^{18}$O=16 was obtained for the cool RCB star SAps; the other RCB
stars in our sample were too warm to display CO molecular lines in their
spectra. Our result $^{16}$O/$^{18}$O=16 for S Aps contrasts with Clayton et al.
's ratio of 4. Our spectra show the gain in information resulting from the
ability at the higher resolution to distinguish clearly  different CO isomers.
High-resolution spectra are essential to derive reliable and precise oxygen
isotopic ratios for these hydrogen-deficient stars.

With S Aps showing a nearly 50-fold difference in the $^{16}$O/$^{18}$O ratio
from the analysable HdC stars, the question arises - is a higher ratio
characteristic of the RCBs? If so, this presumably provides a clue to
understand the evolutionary relationship between HdC and RCB stars. 
Unfortunately, no new information may be drawn about the oxygen isotopic ratio
in HdC stars until more HdC stars are discovered, but there are several RCBs
with CO bands of a suitable strength not yet observed at high resolution. The
example of S Aps shows that RCBs should be reobserved at higher resolution
before drawing conclusions about the HdC-RCB connection. Therefore, we have
extended high-resolution K-band observations to another five cool RCB stars in
order to establish the range of the $^{16}$O/$^{18}$O ratio among RCBs (Clayton
et al.'s range from the 6 RCBs was 1 to $\geq$12).  Section 2 describes the
high-resolution infrared spectroscopic observations and gives an overview of the
observed spectra. Our abundance analysis and the results obtained are presented
in Section 3 and 4, respectively. The derived oxygen isotopic ratios are
discussed in the context of the HdC-RCB connection in Section 5. Final
conclusions are presented in Section 6.

\section{Observations and K-band spectra}

Our new sample contains six cool (T$_{\rm eff}$$\leq$6250 K) RCB stars plus one
HdC star. Five of these stars are observed for the first time at high
resolution. The RCB star S Aps and the HdC star HD 173409 were observed in 
spectral regions not previously covered by us (Garc\'\i a-Hern\'andez et al.
2009). High-resolution (R$\sim$50,000) spectroscopic observations were
obtained in the period February$-$May 2008 with the PHOENIX spectrograph at
Gemini South (Hinkle et al. 2003). We used the 0.34 arcsec slit at  grating
tilts  centered at 2.251, 2.333, 2.349 and 2.365 $\mu$m.
Each one of these wavelength settings provides a spectral coverage of $\sim$0.01
$\mu$m (or 19.5 cm$^{-1}$). 
Exposure times ranging from $\sim$30 seconds (K=5.4)
to $\sim$10 minutes (K=8.3) per wavelength band were needed in order to reach a
high S/N ($>$100 at the continuum). The spectral regions covered for each star
in our sample are listed in Table 1. The observed spectra were reduced to one
dimension by using standard tasks in IRAF\footnote{The Image Reduction and
Analysis Facility software package (IRAF) is distributed by the National Optical
Astronomy Observatories, which is operated by the Association of Universities
for Research in Astronomy, Inc., under cooperative agreement with the National
Science Foundation.} and the telluric features were removed with the help of a
spectrum of a hot star observed the same night. Standard air wavelengths are
given hereafter.

Before attributing an  observed spectrum to the stellar photosphere,  we
determine first if a  star displays a significant infrared excess at 2.3 $\mu$m.
For this purpose,  we constructed spectral energy distributions (SEDs) of the
stars in our sample by using the BVRIJHKL photometry available in the literature
and the IRAS flux density at 12 $\mu$m. In the K band, the flux from RCB stars
SV Sge, S Aps\footnote{According to ASAS All Star Catalogue (ASAS-3)
observations, S Aps was recovering from a weak minimum during our
observations.}, WX CrA and U Aqr is predominantly from the photosphere; an
infrared excess from circumstellar dust does not significantly contaminate our
infrared spectra. The photospheric spectrum is also dominant for the HdC star HD
173409 with an effective temperature of $T_{\rm eff} = 6100$ K. Spectra of V CrA
and ES Aql are contaminated by infrared emission from dust. 

Inspection of the light curve provided by the American Association of Variable
Star Observers (AAVSO)\footnote{see http://www.aavso.org/} and the ASAS All
Star Catalogue (ASAS-3)\footnote{see
http://www.astrouw.edu.pl/asas/?page=aasc}, shows that V CrA experienced a
decline of at least six magnitudes beginning last March 2007. AAVSO measurements
are not reported for the period of our observations but the ASAS-3
measurements confirm that V CrA was at minimum during our observations.
Certainly,  our infrared spectra are featureless. CO and CN lines are expected
to be quite strong in V CrA's photospheric spectrum ($T_{\rm eff} = 6250$ K,
Asplund et al. 2000) and, therefore, the observed featureless spectrum is most
likely from the circumstellar dust.

ES Aql experienced a strong decline in brightness beginning in November 2007. 
Fortunately, ES Aql was recovering from minimum light when observed by us; it
was about one magnitude below maximum light according to AAVSO and ASAS-3
observations. Although ES Aql's photospheric spectrum may be partially diluted 
by emission from circumstellar dust, CO and CN lines are clearly detected in
this star with $T_{\rm eff} = 5000$ K. 

The observed spectral regions were chosen primarily to catch a mix of
$^{12}$C$^{16}$O, $^{12}$C$^{18}$O as well as $^{12}$C$^{14}$N lines. The region
centered at 2.251 $\mu$m was selected in order to be able to estimate the
N abundance from CN lines. The CN molecule (Red System) contributes in all
observed regions but this region is free from contributions from the
first-overtone CO bands. In addition, a few lines of the C$_2$ Phillips bands
0-2 and 1-3 together with some S\,{\sc i} lines cross this interval. Figure 1
shows the observed spectra around 2.251 $\mu$m for the stars in our sample. 

The region centered at 2.333 $\mu$m provides a selection of P and R lines
from the $^{12}$C$^{16}$O 2-0 and 3-1 bands as well as a few weak
$^{12}$C$^{14}$N lines and the Na\,{\sc i} line at 2.3348 $\mu$m. The isomer
$^{12}$C$^{18}$O is not a contributor to this region. The observed spectra in
the latter region are shown in Figure 2. 

Three stars were also observed in the interval around 2.349 $\mu$m. This region
includes the 2-0 $^{12}$C$^{18}$O bandhead and also the 3-1 $^{12}$C$^{17}$O and
4-2 $^{12}$C$^{16}$O bandheads at 2.351 and 2.352 $\mu$m, respectively. Figure 3
shows the 2.349 $\mu$m $^{12}$C$^{18}$O 2-0 band of three cool RCB stars in our
sample in comparison with the $^{18}$O-rich HdC star HD 137613 (Garc\'\i
a-Hern\'andez et al. 2009). Note that strong $^{12}$C$^{18}$O lines are seen in
the spectrum of the HdC star HD 137613, while these lines are weaker in the cool
RCB stars U Aqr, ES Aql and S Aps.  Yet, the 4-2 $^{12}$C$^{16}$O bandhead (and
other $^{12}$C$^{16}$O lines) are stronger in the RCB spectra than in the
spectrum of HD 137613.

Finally, all stars were observed in the region centered at 2.365 $\mu$m. This
region is well suited to providing an estimate of the $^{16}$O/$^{18}$O ratio
because it is longward of the 2-0 $^{12}$C$^{18}$O R-branch head at 2.349 $\mu$m
and includes weak and strong $^{12}$C$^{16}$O and $^{12}$C$^{18}$O lines that
are cleanly separated at R$=50,000$. There are also $^{12}$C$^{14}$N lines in
this region. Figure 4 shows the observed spectra around 2.365 $\mu$m for the
cool RCB stars in our sample. It is clear that the RCB star WX CrA is strongly
enriched in $^{18}$O because lines of   $^{12}$C$^{16}$O and $^{12}$C$^{18}$O
are similar in strength, as anticipated from medium-resolution spectra (Clayton
et al. 2007).  In contrast, the $^{12}$C$^{18}$O lines - especially the cleanest
$^{12}$C$^{18}$O line at $\sim$2.3686 $\mu$m is much weaker in the other four
RCB stars, suggesting that $^{18}$O is less abundant in these stars.

\section{Abundance analysis}

We derive oxygen isotopic abundances  by  using spectral synthesis techniques. A
detailed description of the adopted abundance analysis procedure as well as the
input physics and atomic and molecular linelists has been  presented by us
(Garc\'\i a-Hern\'andez et al. 2009) and will not be repeated here. In short, we
have  constructed H-deficient spherically symmetric MARCS model atmospheres for
the stars in our sample and we considered up-to-date lists of atomic and
molecular lines. The most important contributor of lines in our spectral regions
are the CO and CN molecules. We consider linelists for all the CO and CN isomers
(i.e., $^{12}$C$^{16}$O, $^{12}$C$^{17}$O, $^{12}$C$^{18}$O, $^{12}$C$^{14}$N,
$^{12}$C$^{15}$N, etc.).

The MARCS models follow the prescription discussed by Asplund et al. (1997)
where the input chemical composition is that of the RCB stars (i.e.,
log$\epsilon$(H)=7.5 and a C/He ratio of 1\% by number density\footnote{The
abundances are normalized to log $\sum \mu_i$ $\varepsilon_{i}$ =12.15 where
$\mu_i$ is the mean atomic weight of element $i$, i.e.,
log$\varepsilon$(He)=11.52 and log$\varepsilon$(C)=9.52, see Asplund et al.
(1997).}; Rao \& Lambert 1994). The only exception with respect to initial
composition was U Aqr (see below) for which we have also used a metal-poor model
with [Fe/H]=$-$1.0 dex. The TURBOSPECTRUM package (Alvarez \& Plez 1998), which
is compatible with MARCS, is used to generate the corresponding synthetic
spectra. 

Apart from the adopted input composition, a microturbulent velocity, surface
gravity and effective temperature must be chosen. The warm RCB stars display an
average microturbulent velocity of 6.6 km s$^{-1}$ with no apparent trend with
effective temperature (Asplund et al. 2000). Thus, we assumed a microturbulence
of 7 km s$^{-1}$ for all RCB stars.  This value provides a fair fit to both the
weaker and strong lines of a given species, a necessary condition according to
the definition of microturbulence. When a synthetic spectrum computed for a
microturbulence of 7 km s$^{-1}$ is convolved with a  Gaussian profile with a
width of 6 km s$^{-1}$ in order to take into account the instrumental profile at
R$=$50,000, a good fit is obtained to the line profiles. This suggests that the
macroturbulence is small. In addition, a surface gravity of $\log$g$=+0.5$ was
assumed in the construction of the model atmospheres for all RCB stars. This is
consistent with the gravities found by Asplund et al. (2000) in their optical
spectroscopic analysis of the cooler ( $T_{\rm eff}$$<$6500 K) RCB stars.
Finally, for each star in our sample we chose the effective temperatures listed
by Clayton et al. (2007) for their analysis of medium resolution near-infrared
spectra (see their Table 2). The adopted $T_{\rm eff}$  are listed in Table 2.
It is to be noted that we do not precisely know the effective temperatures of
our sample stars and they cannot be derived from our high-resolution spectra.
Note also that by selecting the Clayton et al's temperatures we can compare our
derived oxygen isotopic ratios with those obtained from medium-resolution
spectra. In any case, our main goal is to determine the $^{16}$O/$^{18}$O oxygen
isotopic ratios, and their determination is almost independent of the adopted
model parameters, as shown in our previous paper and in Table 2 for SV Sge and U
Aqr for which two different models are used for each star. According to our
near-infrared and our previous optical spectra SV Sge looks hotter than the 4000
K  adopted by Clayton et al. U Aqr according to Bond et al. (1979) from their
analysis of a few strong optical lines has a calcium abundance similar to that
of HD 182040, i.e., the metallicity is approximately solar. (U Aqr is noted for
its remarkably high abundances of light $s$-process abundances  - Sr and Zr.)
Clayton et al. (2007) adopted $T_{\rm eff} = 6000$ K from Asplund et al. (1997).
Cottrell \& Lawson (1998) suggest U Aqr belongs to the Galactic halo. Then, it
might be metal-poor. There is spectroscopic evidence for this and for a lower
effective temperature (Rao 2008). Thus, for U Aqr we consider a model with
$T_{\rm eff} = 6000$ K and a metal-poor ([Fe/H] = -1) model with $T_{\rm eff} =
5400$ K.

The corresponding synthetic spectra were compared to the observed ones in order
to determine the isotopic ratios for O and N - see Garc\'\i a-Hern\'andez et al.
(2009) for a detailed description of the procedure employed in the derivation of
isotopic abundances in hydrogen-deficient stars. We also list the derived O and
N elemental abundances for the assumed input composition (i.e.,
log$\varepsilon$(H)=7.5, log$\varepsilon$(He)=11.52 and
log$\varepsilon$(C)=9.52) together with, when possible, the abundances of the
heavier elements Na and S from the Na\,{\sc i} line at 2.3348
$\mu$m\footnote{Note that the 2.3348 $\mu$m Na\,{\sc i} line is strongly blended
with molecular features, preventing us of measuring a reliable Na abundance in
most of the sample stars.} and the few S\,{\sc i} lines around 2.251 $\mu$m. It
should be noted here that the listed O, N, Na, and S elemental abundances must
be considered with caution because they are strongly dependent of the
atmospheric parameters (Table 2). The formal error in the total N and O
abundances taking into account variations of the atmospheric parameters used in
the modeling of hydrogen-deficient stars are estimated to be of the order of
0.3$-$0.4 dex (Garc\'\i a-Hern\'andez et al. 2009). In addition, the C$_2$
Phillips system lines 0-2 Q(34), 1-3 Q(8) and 1-3 Q(10) (see Fig.1) observed in
HdC and cool RCB stars exhibit a `carbon problem', which complicates their use
for carbon abundance determinations (Garc\'\i a-Hern\'andez et al. 2009 and
references therein). Thus, we have to assume the input C abundance
(log$\varepsilon$(C)=9.52) in our abundance analysis, which affects the derived
N and O elemental abundances. The uncertainties for the derived Na and S
abundances are even higher than those estimated for N and O with a typical value
of $\sim$0.6$-$0.7 dex - especially in these cool RCB stars where the atomic
lines are usually blended with the strong CN and CO molecular features across
the spectra.

\section{Oxygen isotopic ratios}

For the RCB stars WX CrA, ES Aql, U Aqr, and SV Sge we are able for the first
time to get reliable estimates of the $^{16}$O/$^{18}$O isotopic ratio. The
$^{16}$O/$^{18}$O ratio of 16 obtained for the RCB star S Aps was previously
known from spectrum synthesis at 2.349 $\mu$m (Garc\'\i a-Hern\'andez et al.
2009). For the RCB star V CrA, the K-band spectra are featureless because they
are dominated by emission from circumstellar dust (see Sect. 2) precluding a
measurement of the $^{16}$O/$^{18}$O ratio. For the five RCB stars, lower limits
to the $^{16}$O/$^{17}$O and $^{14}$N/$^{15}$N ratios from the absence of
detectable $^{12}$C$^{17}$O and $^{12}$C$^{15}$N lines are also determined.
Unfortunately, we can not give estimates of the $^{12}$C/$^{13}$C ratios because
our observations did not cover the 2-0 $^{13}$C$^{16}$O bandhead at 2.344
$\mu$m. Individual $^{13}$C$^{16}$O features are not detectable but this is not
surprising given that previous albeit scanty evidence for RCBs is that they have
a high $^{12}$C/$^{13}$C ratio. Note that we observed the HdC star HD 173409
around 2.251 $\mu$m and a lower limit to the $^{14}$N/$^{15}$N ratio as well as
the S elemental abundance from the S\,{\sc i} lines are also given. Table 2
summarises the derived abundances. 

By inspection of Table 2,  the $^{16}$O/$^{18}$O ratios (with the sole exception
of WX CrA) of these RCB stars  differ from those reported by Clayton et al.
(2007) from medium-resolution spectra.  Clayton et al. (2007) underestimated the
$^{16}$O/$^{18}$O ratios for the RCB stars ES Aql, S Aps and SV Sge. This may be
because most of the $^{12}$C$^{16}$O lines in these three stars are saturated,
not being useful for abundance analysis - we derive the isotopic $^{16}$O and
$^{18}$O abundances from the weaker and cleanest lines in the observed spectra
around 2.333 and 2.365 $\mu$m. This is illustrated in Figure 5 for the RCB star
SV Sge. The $^{16}$O/$^{18}$O ratio of 16 is obtained in this star for which
Clayton et al. obtained a value of 4. For those stars (e.g., ES Aql, U Aqr and S
Aps) with available spectra in the 2.349 and 2.365 $\mu$m regions, the derived
$^{16}$O/$^{18}$O ratios from both regions are in excellent agreement. For the
RCB star U Aqr, our $^{16}$O/$^{18}$O estimate of 4 contrasts with the lower
limit of 12 estimated by Clayton et al. Synthetic spectra  for U Aqr around
2.333 and 2.365 $\mu$m are shown in Figures 6 and 7, respectively.  The
$^{16}$O/$^{18}$O ratio of 20 displayed by  ES Aql contrasts with Clayton et
al.'s estimate of 6 for this star. Figures 8 and 9 show synthetic spectra  for
this star around the 2-0 $^{12}$C$^{18}$O bandhead at 2.349 $\mu$m and around
2.365 $\mu$m, respectively. It is to be noted that the $^{12}$C$^{16}$O lines
are saturated for this star with  $T_{\rm eff}$$=$5000 K. In addition, the
K-band  spectra for ES Aql are partially diluted by circumstellar dust with
dilution decreasing between 2008 March and May. This is deduced from the fact
that synthetic spectra predict stronger $^{12}$C$^{16}$O lines than observed and
the CO lines are stronger in the 2008 May spectrum than in 2008 April. 
ASAS-3 visual observations show that the star have brightened by about one
magnitude between March and May when it was about 0.5 magnitude below maximum
light. Although the derived $^{16}$O and $^{18}$O isotopic abundances differ by
about 0.2 dex from analyses of the 2008 March 21 and April 8 spectra, the
$^{16}$O/$^{18}$O ratio of 20 remains unchanged (Figures 8 and 9). Finally, the
RCB star WX CrA seems to be a HdC-like ($^{18}$O-rich) star with
$^{16}$O/$^{18}$O=0.3 where a value of 1 was provided by Clayton and colleagues.
Synthetic and observed spectra around 2.365 $\mu$m for WX CrA are shown in
Figure 10.

The $^{16}$O/$^{18}$O ratios derived from high-resolution spectra are in fair
agreement with those derived from medium-resolution spectra  when the molecular
lines of the different CO isomers are not strongly saturated. This is the case
for the $^{18}$O-rich HdC stars (Garc\'\i a-Hern\'andez et al. 2009) and the RCB
star WX CrA. In short, high-resolution spectra are needed in order to derive 
reliable oxygen isotopic ratios in cool RCB stars with strong CO molecular
lines.

In summary, there is a clear difference between the $^{16}$O/$^{18}$O ratios of
the HdC and RCBs for those stars for which a measurement is possible. This
difference suggested by interpretation of medium-resolution spectra is confirmed
and extended by our high-resolution spectra. All (three!) HdCs with CO lines
show $^{18}$O more abundant than $^{16}$O. All but one of the investigated RCBs
shows $^{16}$O at least ten times more abundant than $^{18}$O.\footnote{The cool
RCB Z UMi has $^{16}$O/$^{18}$O $\geq 8$ according to Clayton et al. (2007).}

\section{The HdC-RCB connection}

Given the extreme rarity of HdC and RCB stars, it seems reasonable to suppose
that they are closely related.  Here, we adopt the view that both are products
of the DD scenario in which a He white dwarf merges with a C-O white dwarf. The
merger inflates the resultant atmosphere around the C-O white dwarf to
supergiant dimensions. A hot H-deficient supergiant is created that is expected
to  evolve rapidly to lower temperatures at approximately constant luminosity.
The star is a cool supergiant for about  10$^5$ -- 10$^6$ years before the onset
of  rapid evolution back to high temperatures and a final descent of the white
dwarf cooling track. HdCs and RCBs are to be identified with the cool supergiant
phase. 

The precise form of the evolutionary track in the H-R diagram may depend on
multiple aspects of the merger, e.g.,  the masses and compositions of the white
dwarfs, the rate of mass transfer from the He white dwarf to the C-O white
dwarf,  and the gravitational and nuclear energy release at the C-O white dwarf.
Yet, it is very likely that a  supergiant product of a merger spends  most of
its time as a cool star. If a HdC  results, the evolutionary sequence will be
HdC $\rightarrow$ RCB $\rightarrow$ EHe. Perhaps, some mergers do not populate
the coolest parts of the supergiant track, then the sequence will  be RCB
$\rightarrow$ EHe. In other words, all HdC may evolve to RCB and EHe stars, but,
perhaps, not all RCB and EHe stars were HdC stars. Here, we examine the
abundances in light of the assumption that   HdC $\rightarrow$ RCB $\rightarrow$
EHe is the common sequence. 

It is assumed for the purposes of this discussion that the remarkable high
concentration of $^{18}$O is created during or shortly after the merger.
Although it is plausible that the $^{18}$O is synthesized from $^{14}$N (Warner
1967), no explanation has yet been provided as to why the $^{16}$O/$^{18}$O
ratio takes the low values seen in the HdCs and why surviving $^{14}$N  seems so
slightly depleted. These questions were explored by Clayton et al. (2007) and
Garc\'{\i}a-Hern\'{a}ndez et al. (2009). Here, we set such matters aside and 
discuss how the difference in $^{16}$O/$^{18}$O ratios between HdC and RCB stars
might be explained.

Abundances derived from the infrared spectra for the HdC and cool RCBs  betray
some intriguing differences between the two groups of H-deficient stars. Table 3
gives the N, O and $^{16}$O and $^{18}$O, Na, and S abundances for stars from
this and our previous paper. The C abundance is assumed to be the same for all
stars (see Table 2). These differences are unlikely to be attributable to
systematic differences in atmospheric parameters between the HdC and RCB stars;
the differences on average are slight except, perhaps, for the C/He ratio which
is an assumed not a derived parameter (see below for our conjecture about this
ratio).

The most striking difference between HdC and RCB stars is seen among the
estimates of the $^{16}$O/$^{18}$O ratios. The three HdCs for which the ratio is
measureable have an isotopic ratio near 0.5. Three of the four RCBs with a
measured isotopic ratio have a ratio at least one order of magnitude higher than
that seen in the HdCs. The exception is the  RCB  WX CrA which has an HdC-like
ratio. It bears repeating (see Table 2) that the measured $^{16}$O/$^{18}$O
ratio is insensitive to the adopted model atmosphere over quite wide variations
of the model atmosphere parameters. Furthermore, this HdC-RCB difference  is
evident from inspection of the spectra. 

In order to interpret this clear difference in the $^{16}$O/$^{18}$O ratio 
between HdC and RCB stars it is helpful to know the contributions from an
increase in the $^{16}$O and a decrease in the $^{18}$O abundance between the
two groups of stars. The $^{16}$O and $^{18}$O abundances are given in Table 3
and shown in Figure 11. One interpretation of the abundances is that the
difference in $^{16}$O/$^{18}$O ratios may be partially attributable to  
decrease in $^{18}$O abundances between  the $^{18}$O-rich HdC stars (and WX
CrA) but primarily arises from an increase in the $^{16}$O abundance between HdC
and RCB stars.  (The lower than average  $^{16}$O and $^{18}$O abundances for ES
Aql among the RCBs may possibly be due to the uncertain and varying dilution of
the spectrum by dust emission.)
 
Here, we do not attempt to explain the origin of the $^{18}$O that is such a
remarkable characteristic of the HdC - see Clayton et al. (2007) and our 2009
paper for speculations. In this section, we directly speculate to how the
differences in composition between HdC and RCB stars may be related to the
dredge-up of nuclear-processed material by the supergiant as it evolves from a
HdC to a RCB star. Published calculations  bounding the possibilities have yet
to be reported. Our speculations draw on the few models presented by Saio \&
Jeffery (2002) for mergers in which mass transfer is a slow process continuing
well into the supergiant phase. In these models, the convective envelope just
prior to the rapid evolution back to high temperatures makes a close approach to
the He-burning shell. (This close approach may  not occur in the case of a
merger which results in traces of hydrogen in the accreted material  and then 
the maintenance of a H-burning shell in the supergiant.)

With a close approach to the He-burning shell, the $^{16}$O  in the envelope may
increase. Such an increase is more certain to occur, if after the cessation of
He-burning,  the convective envelope is  able to penetrate to the
He-shell/CO-core boundary. Also, the $^{18}$O abundance is likely to decrease
because it may have been destroyed in the deepest layers; $^{18}$O is more prone
to $(\alpha,\gamma)$ destruction than $^{16}$O by several orders of magnitude. A
few per cent addition of mass from the He-exhausted layers should suffice to
increase the $^{16}$O abundance by an order of magnitude. 

In this scenario, the $^{12}$C abundance of the envelope and atmosphere
increases too between HdC and RCB stars because the ratio of $^{12}$C/$^{16}$O
in the He-exhausted layers is most likely to be greater than one.  The He
abundance should not be significantly changed by the extension of the deep
convective envelope  across the He-shell/C-O-core boundary. Thus, the RCBs may
have a higher C/He ratio than the HdCs. This may account also for an intriguing
difference in Na and S abundances between HdC and RCB stars in Table 3. Both S
and Na are about 0.6 dex higher in HdCs than in the RCBs. 

Derived abundances are dependent on the assumed C/He ratio. In the infrared
He$^-$ free-free opacity is the leading contributor of continuous opacity with C
as an important electron donor.  In our previous paper, we reported that a
change in the C/He ratio from 1\% to 10\% lead to an increase of Na and S
abundances by about 0.6 dex. This suggests that decreasing the C/He ratio to
about 0.1\% will reduce the Na and S abundances to about their levels seen in
the RCB stars. Derived abundances for N and O change little with a change in
C/He ratio. The C/He ratio of 1\% for the HdCs and RCBs is an assumed value. A
change of the C/He ratio from 0.1\% to 1\% to reconcile Na and S abundances in
HdC with values for RCBs is possibly consistent with the C/O ratio of material
dredged-up from the He-exhausted layers.

Although this scenario offers a scheme for accounting for the lowering of the
$^{16}$O/$^{18}$O ratio between HdC and RCB stars and also for the difference in
their abundances of Na and S, it augments the  challenge in accounting for the
compositions of the warm RCBs (and EHes) for which detailed results are
available (Asplund et al. 2000; Pandey et al. 2006).  If the RCBs are affected
by a late dredge-up, one would suppose that the final C/He must vary from star
to star on account of the different amounts of mass dredged up and the different
C/O ratios. Such a variation would be reflected in star-to-star variations in
abundances when all stars are analysed assuming the same C/He ratio.  Yet, the
abundances of elements heavier than O are quite similar star-to-star across the
sample of majority RCBs.  The dispersion including the measurement uncertainties
is only about 0.2 dex for well observed elements over the sample of 14 warm
RCBs. 

The challenge in accounting for the uniform compositions was not created by
incorporating a late dredge-up into the evolution of these H-deficient
supergiants. Its origin surely lies in the DD scenario itself. Mergers of He
white dwarfs with C-O white dwarfs spanning a range in masses and previous
histories of the white dwarfs might be expected to result in H-deficient stars
with quite different C/He ratios and metallicities. 

One looks forward to detailed modelling of H-deficient supergiants that will
test our conjecture about a late dredge-up of material from He-exhausted layers.

\section{Concluding Remarks}

Challenges are not a new phenomenon to students of the R Coronae Borealis stars.
High-resolution infared spectroscopy promises to provide responses to some of
the key challenges regarding the RCBs and their putative relatives the HdC and
EHe stars. Here, we suggested an evolutionary link between the HdCs and the RCBs
involving the dredge-up of material to the surface. Further exploration of the
link is possible. Unfortunately, the list of known HdCs has been exhausted. A
few cool RCBs remain to be observed in the K band at high resolution. Of these,
the most critical is V CrA. Unfortunately, our observations were obtained when
the star was at minimum and infrared emission came from the dust shell and not
the stellar photosphere. V CrA is special target because it is one of four known
minority RCB stars (Rao \& Lambert 1994) and the only one likely to show CO
lines. Minority RCBs are distinguished by their high S/Fe and Si/Fe ratios
(among other abundance anomalies): V CrA has [Si/Fe] $\simeq$ [S/Fe] $\simeq$ 2
(Asplund et al. 2000; Rao \& Lambert 2008). Additionally, V CrA is rich in
$^{13}$C with $^{12}$C/$^{13}$C $\simeq 3$ (Rao \& Lambert 2008). Among those
HdC and RCBs for which the carbon isotopic ratio may be estimated,  V CrA is the
only known example with a high $^{13}$C abundance. What does the combination of
anomalous abundance ratios and high $^{13}$C abundance imply about the origin 
of V CrA and minority RCBs? Is their origin a variant of the DD scenario or must
another (i.e., the FF) scenario be invoked?  Measurement of the oxygen isotopic
ratios is  potentially a valuable clue to the answers to these questions. 

Infrared spectroscopy offers the possibility of probing the carbon problems
posed by the RCBs and HdCs - see Asplund et al. (2000) and
Garc\'{\i}a-Hern\'{a}ndez et al. (2009) - in the failure to account for the
strengths of the atomic and molecular carbon lines. From the optical to the
infrared, the leading contributor to the continuous opacity depending on the
star's effective temperature may change from the photoionization of neutral
carbon to free-free absorption from neutral helium. Full spectral coverage at
high spectral resolution from the optical through at least the K band offers a
novel opportunity to probe and hopefully to resolve the carbon problems. With
their resolution, the chemical compositions of these fascinating H-deficient
stars should be on a firmer footing and  insights gained into their origins.

\acknowledgments{D.A.G.H. acknowledges support for this work provided by the
Spanish Ministry of Science and Innovation (MICINN) under the 2008 Juan de La
Cierva Program and under grant AYA-2007-64748. D.A.G.H. also acknowledges the
great hospitality of Dr. N. Kameswara Rao during his stay at the Indian
Institute of Astrophysics (Bangalore, India). This paper is based on
observations obtained at the Gemini Observatory, which is operated by the
Association of Universities for Research in Astronomy, Inc., under a cooperative
agreement with the NSF on behalf of the Gemini partnership: the National Science
Foundation (United States), the Science and Technology Facilities Council
(United Kingdom), the National Research Council (Canada), CONICYT (Chile), the
Australian Research Council (Australia), Minist\'{e}rio da Ci\^{e}ncia e
Tecnologia (Brazil) and Ministerio de Ciencia, Tecnolog\'\i a e Innovaci\'on
Productiva (Argentina). The observations were obtained with the Phoenix infrared
spectrograph developed by the National Optical Astronomy Observatory. The
spectra were obtained as part of program GS-2008A-Q-11. This research has been
supported in part by grant F-634 to D.L.L. from the Robert A. Welch Foundation
of Houston, Texas. K.E. gratefully acknowledges support from the Swedish
Research Council.}

\clearpage

\begin{deluxetable}{lcc}
\tabletypesize{\scriptsize}
\tablecaption{Observation summary of the near-IR PHOENIX observations$^{a}$\label{tbl-1}}
\tablewidth{0pt}
\tablehead{
\colhead{Star} & \colhead{Spectral ranges observed}  & \colhead{Date}  \\
\colhead{} & \colhead{$\mu$m}  & \colhead{}  \\
}
\startdata
ES Aql         & 2.251, 2.332, 2.349, 2.366 &  2008 Apr 8, 2008 Mar 21, 2008 May 13   \\
SV Sge         & 2.251, 2.332, 2.366        &  2008 May 13  \\
S Aps          & 2.332, 2.366               &  2008 Feb 10   \\ 
WX CrA         & 2.332, 2.366               &  2008 Apr 10, 2008 Mar 20  \\ 
U Aqr          & 2.251, 2.332, 2.349, 2.366 &  2008 May 13   \\
V CrA          & 2.251, 2.332, 2.349, 2.366 &  2008 Feb 29, 2008 Mar 20, 2008 Apr 7$-$10 \\  
HD 173409      & 2.251                      &  2008 Mar 21   \\
\enddata

\tablenotetext{a}{The first six objects are cool RCB stars
while HD 173409 is a HdC star.}
\end{deluxetable}

\begin{deluxetable}{lcccccccc}
\tabletypesize{\scriptsize}
\tablecaption{Derived chemical abundances in cool RCB stars$^{a}$\label{tbl-2}}
\tablewidth{0pt}
\tablehead{
\colhead{Star} &\colhead{T$_{\rm eff}$ (K)$^{b}$}
&\colhead{$^{16}$O/$^{18}$O$^{c}$} &
\colhead{$^{16}$O/$^{18}$O}  & \colhead{$^{16}$O/$^{17}$O} &
\colhead{$^{14}$N/$^{15}$N}  & \colhead{C/N/O$^{d}$} &
\colhead{ Na} & \colhead{$<$S$>$} \\
}
\startdata
ES Aql$^{e}$ &5000 & 6        & 20$\pm$5    & $>$50  & $>$10   & 9.5/8.6/8.6       &$\dots$         & 6.8      \\
SV Sge       &4000 & 4        & 16$\pm$4    & $>$20  & $>$10   & 9.5/7.7/8.4       &$\dots$         & 7.6      \\
$\dots$      &5000 & $\dots$  & 13$\pm$4    & $>$50  & $>$10   & 9.5/9.2/9.0       &$\dots$         & 7.4      \\
S Aps        &5400 & 4        & 16$\pm$4    & $>$80  & $>$10   & 9.5/9.2/9.2       &$\dots$         & 6.8      \\
WX CrA       &5300 & 1        & 0.3$\pm$0.1 & $>$6   & $\dots$ & 9.5/9.3/8.7       & 7.3            & $\dots$  \\
U Aqr        &6000 & $\geq$12 & 4$\pm$1     & $>$50  & $>$10   & 9.5/9.4/9.3       &$\leq$5.4       & 6.5      \\ 
z=$-$1.0     &5400 & $\dots$  & 3$\pm$1     & $>$20  & $>$8    & 9.5/6.8/6.5       &$\leq$5.2       & 6.3      \\ 
V CrA$^{f}$  &6250 & $\dots$  &$\dots$      &$\dots$ &$\dots$  &$\dots$            &$\dots$         & $\dots$  \\     
HD 173409    &6100 & $\dots$  &$\dots$      &$\dots$ & $>$4    & 9.5/9.2/8.7       & 7.0            & 7.6      \\
\enddata
\tablenotetext{a}{The abundances are normalized to log $\sum \mu_i
\epsilon_{i}$=12.15, i.e. log$\varepsilon$(He)=11.52 and
log$\varepsilon$(C)=9.52. The error in the derived $^{16}$O/$^{18}$O ratios are
estimated from the typical uncertainty of $\pm$0.05 dex associated with the
corresponding $^{16}$O and $^{18}$O isotopic abundances.} 
\tablenotetext{b}{References.- Clayton et al. (2007); Asplund et al. (1997,
1998); Lawson et al. (1990).}
\tablenotetext{c}{$^{16}$O/$^{18}$O ratios derived from medium-resolution
near-IR spectra by Clayton et al. (2007).} 
\tablenotetext{d}{The CNO abundances are given for the assumed input composition
i.e. log$\varepsilon$(Fe)=7.2, log$\varepsilon$(Na)=6.8,
log$\varepsilon$(S)=7.5, etc. Asplund et al. 1997)}
\tablenotetext{e}{The K-band spectrum is significantly obscured by circumstellar
dust. Note that the CNO abundances from the simultenous spectra taken on April 8
2008 are given.}
\tablenotetext{f}{The K-band spectrum is largely from circumstellar dust and the
photospheric spectrum is greatly obscured or diluted.} 
\end{deluxetable}

\begin{deluxetable}{lcccccc}
\tabletypesize{\scriptsize}
\tablecaption{Chemical abundances in HdC and cool RCB stars$^{a}$\label{tbl-3}}
\tablewidth{0pt}
\tablehead{
\colhead{Star} &\colhead{log$\varepsilon$(N)} &\colhead{log$\varepsilon$(O)} &
\colhead{log$\varepsilon$($^{16}$O)}  & \colhead{log$\varepsilon$($^{18}$O)}  & 
\colhead{log$\varepsilon$(Na)} & \colhead{log$\varepsilon$(S)} \\
}
\startdata
ES Aql      & 8.6 & 8.6   & 8.6 & 7.3 &$\dots$   & 6.8 \\
SV Sge      & 7.7 & 8.4   & 8.4 & 7.2 &$\dots$   & 7.6 \\
S Aps       & 9.2 & 9.2   & 9.2 & 8.0 &$\dots$   & 6.8 \\
WX CrA      & 9.3 & 8.7   & 8.1 & 8.6 & 7.3	 & $\dots$ \\
U Aqr       & 9.4 & 9.3   & 9.2 & 8.6 &$\leq$5.4 & 6.5 \\ 
HD 137613   & 9.4 & 8.7   & 8.3 & 8.6 &6.9       & 7.6 \\
HD 175893   & 9.2 & 8.7   & 8.1 & 8.6 &6.5       & 7.6 \\
HD 182020   & 9.2 & 8.0   & 7.6 & 7.9 &6.8       & 7.4 \\
\enddata
\tablenotetext{a}{The first five objects are the cool RCB stars analyzed in this
paper while the last three ones are the HdC stars studied by Garc\'\i a-Hern\'andez et al. (2009). The abundances are normalized to log $\sum \mu_i
\epsilon_{i}$=12.15, i.e. log$\varepsilon$(He)=11.52 and
log$\varepsilon$(C)=9.52.} 
\end{deluxetable}

\clearpage

\begin{figure}
\includegraphics[angle=-90,scale=.60]{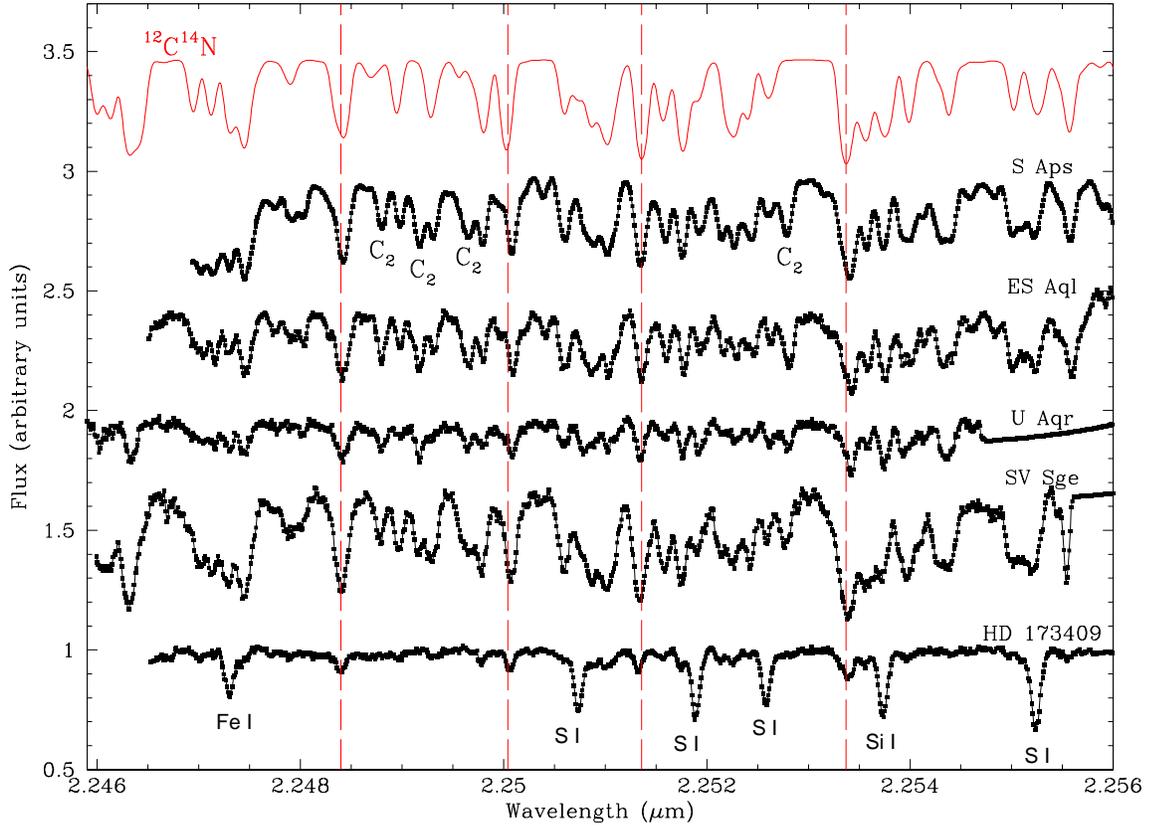}
\caption{K-band spectra centered at $\sim$2.251 $\mu$m for four cool RCBs (from
top to bottom: S Aps, ES Aql, U Aqr, and SV Sge, respectively). The spectrum of
the HdC star HD 173409 is also shown for comparison. Positions of several 
atomic lines  are labeled. The strongest $^{12}$C$^{14}$N features are indicated
with a dashed red vertical line. A $^{12}$C$^{14}$N synthetic spectrum (in red)
composed for S Aps is also shown. Phillips system C$_2$ lines labeled
are 1-3 R22, 0-2 Q34, 1-3 Q8 and 1-3 Q10 in order of increasing wavelength.
\label{fig1}} \end{figure}

\clearpage

\begin{figure}
\includegraphics[angle=-90,scale=.60]{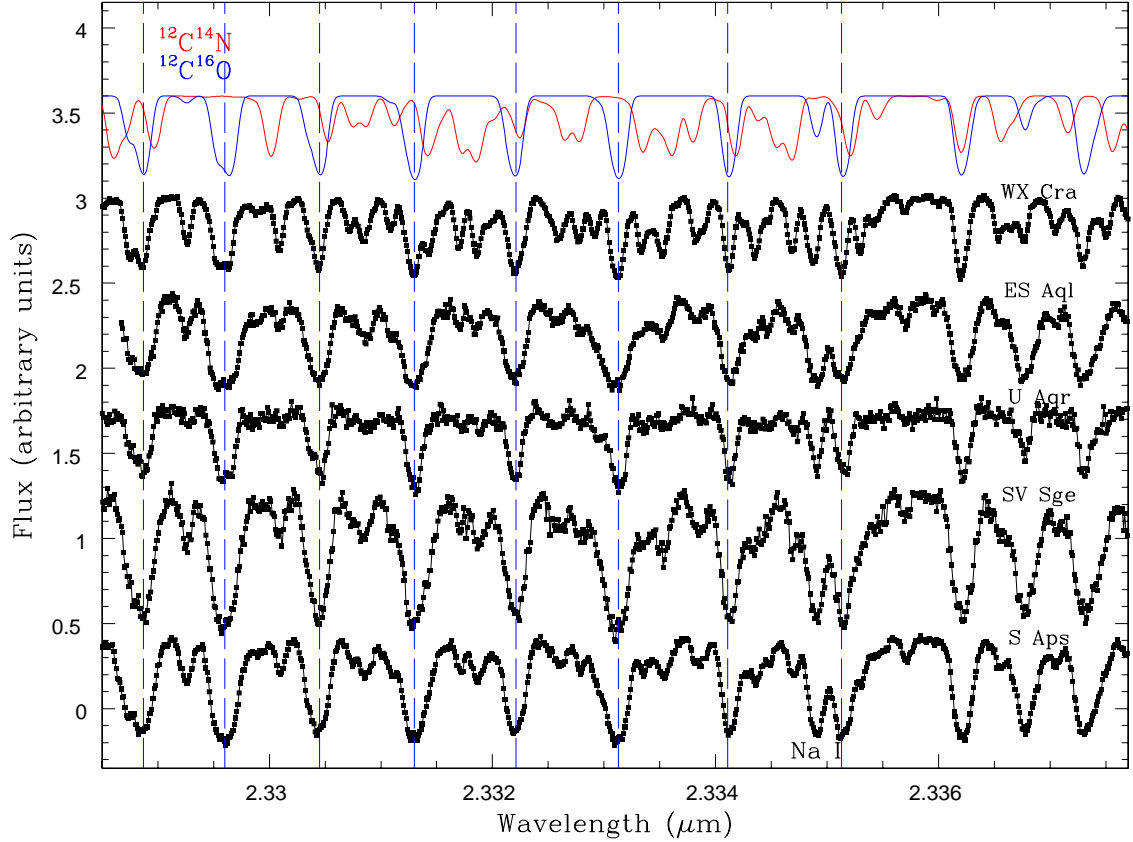}
\caption{K-band spectra centered at 2.333 $\mu$m for the five cool RCB stars in
our sample. This spectral region is dominated by lines of the $^{12}$C$^{16}$O
and $^{12}$C$^{14}$N molecules. Note the Na\,{\sc i} line at 2.3348 $\mu$m that
is usually blended with  molecular features. The strongest $^{12}$C$^{16}$O
lines are marked with a dashed blue vertical line. Synthetic spectra composed
for WX CrA for the CO and CN isomers ($^{12}$C$^{16}$O in blue and
$^{12}$C$^{14}$N in red) indicate that these molecules dominate the spectra.
\label{fig2}} 
\end{figure}

\clearpage

\begin{figure}
\includegraphics[angle=-90,scale=.60]{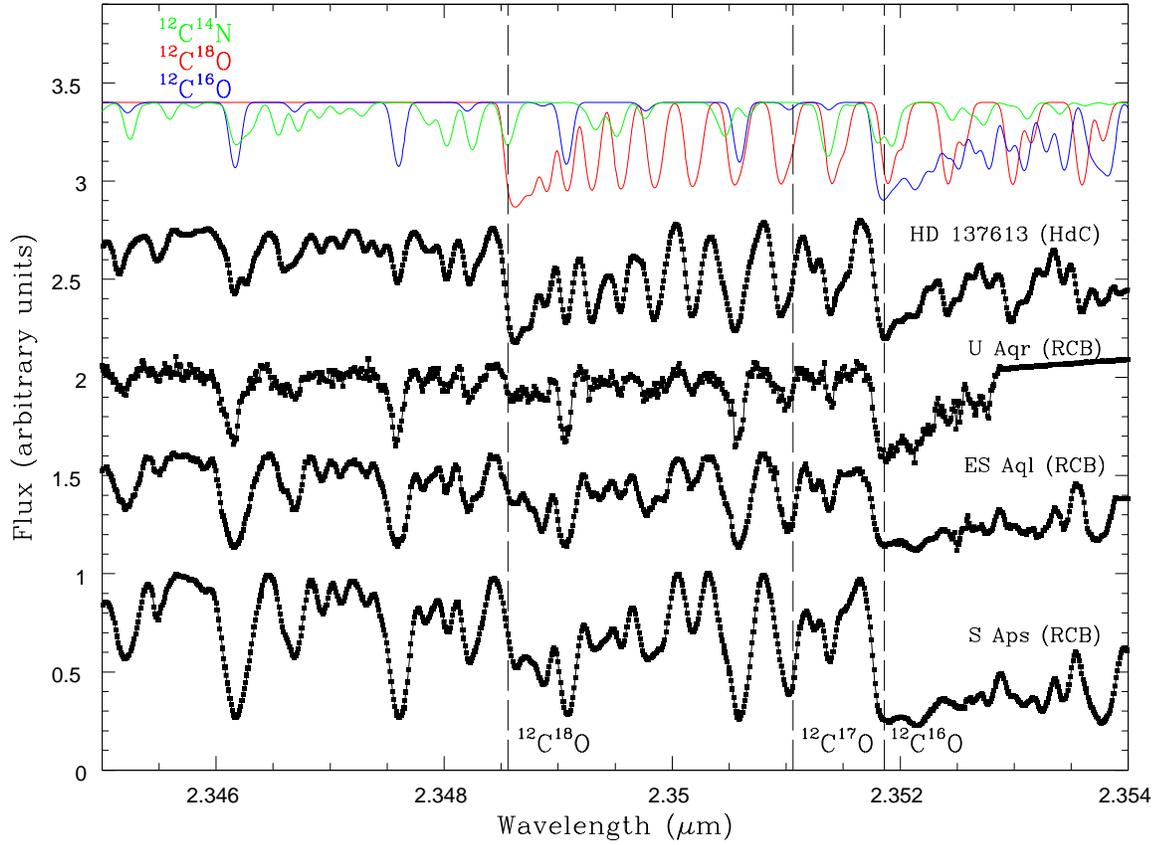}
\caption{K-band spectra centered at $\sim$2.349 $\mu$m for three cool RCB stars 
(U Aqr, ES Aql, and S Aps) and the  $^{18}$O-rich HdC star HD 137613 (taken from
Garc\'\i a-Hern\'andez et al. 2009). Wavelengths of 2-0 $^{12}$C$^{18}$O, 3-1
$^{12}$C$^{17}$O, and the 4-2 $^{12}$C$^{16}$O bandheads are marked with a
vertical dashed line. Synthetic spectra composed for HD 137613 are shown for the
isomers $^{12}$C$^{16}$O in blue, $^{12}$C$^{18}$O in red and $^{12}$C$^{14}$N
in green. \label{fig3}}
\end{figure}

\clearpage

\begin{figure}
\includegraphics[angle=-90,scale=.60]{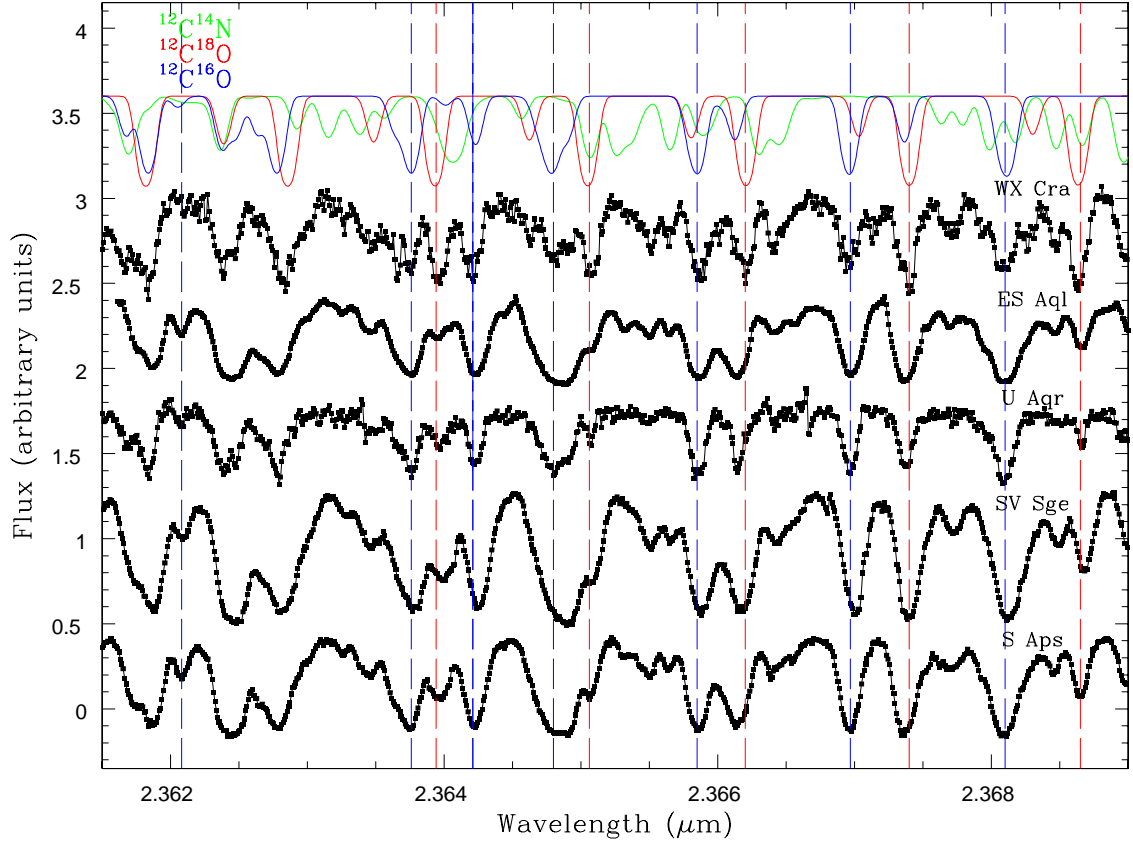}
\caption{K-band spectra centered at 2.365 $\mu$m of the five cool RCB stars in
our sample. Synthetic spectra composed for WX Cra for the CO isomers
($^{12}$C$^{16}$O in blue and $^{12}$C$^{18}$O in red) and for the
$^{12}$C$^{14}$N isomer (in green) are shown at the top for comparison. Some
lines of $^{12}$C$^{18}$O and $^{12}$C$^{16}$O are resolved and marked with
dashed vertical lines. Note that the strongest $^{12}$C$^{16}$O and
$^{12}$C$^{18}$O lines are saturated in the RCB stars ES Aql, SV Sge, and S Aps.
However,  weaker and clean $^{12}$C$^{18}$O and $^{12}$C$^{16}$O lines such as
those at $\sim$2.3686 $\mu$m and $\sim$2.3620 $\mu$m indicate that $^{18}$O is
less abundant in these three RCBs compared with the $^{18}$O-rich RCB WX CrA.
\label{fig4}} 
\end{figure}

\clearpage

\begin{figure}
\includegraphics[angle=-90,scale=.60]{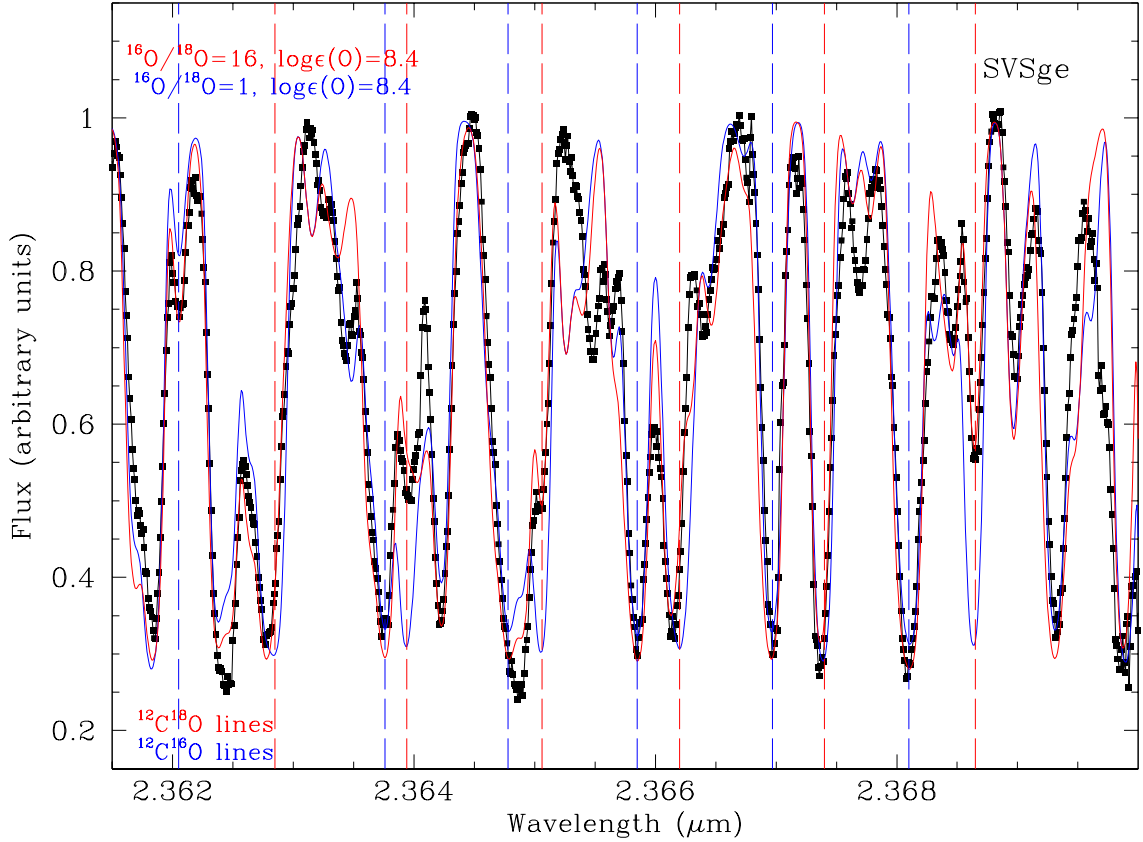}
\caption{Best synthetic (red) and observed (black) spectrum for the region
centered at 2.365 $\mu$m for the RCB star SV Sge. The red spectrum assumes
T$_{\rm eff}$=4000 K, the input C abundance of 9.52, the N abundance of 7.7,  a
total O ($^{16}$O $+$ $^{18}$O) of log$\varepsilon$(O)=8.4 and the isotopic
ratio $^{16}$O/$^{18}$O=16. Selected $^{12}$C$^{18}$O and $^{12}$C$^{16}$O lines
are marked with red and blue vertical dashed lines, respectively.  A synthetic
spectrum (blue) for  the same  total O ($^{16}$O $+$ $^{18}$O) of
log$\varepsilon$(O)=8.4  but a lower $^{16}$O/$^{18}$O ratio of 1 (i.e., a
higher $^{18}$O abundance)   is shown for comparison. Note that the strongest
$^{12}$C$^{16}$O and $^{12}$C$^{18}$O lines are saturated. The weaker and
cleanest $^{12}$C$^{18}$O and $^{12}$C$^{16}$O lines such as those at
$\sim$2.3686 $\mu$m and $\sim$2.3620 $\mu$m, respectively, provide the ratio
$^{16}$O/$^{18}$O=16. \label{fig5}}
\end{figure}

\clearpage

\begin{figure}
\includegraphics[angle=-90,scale=.60]{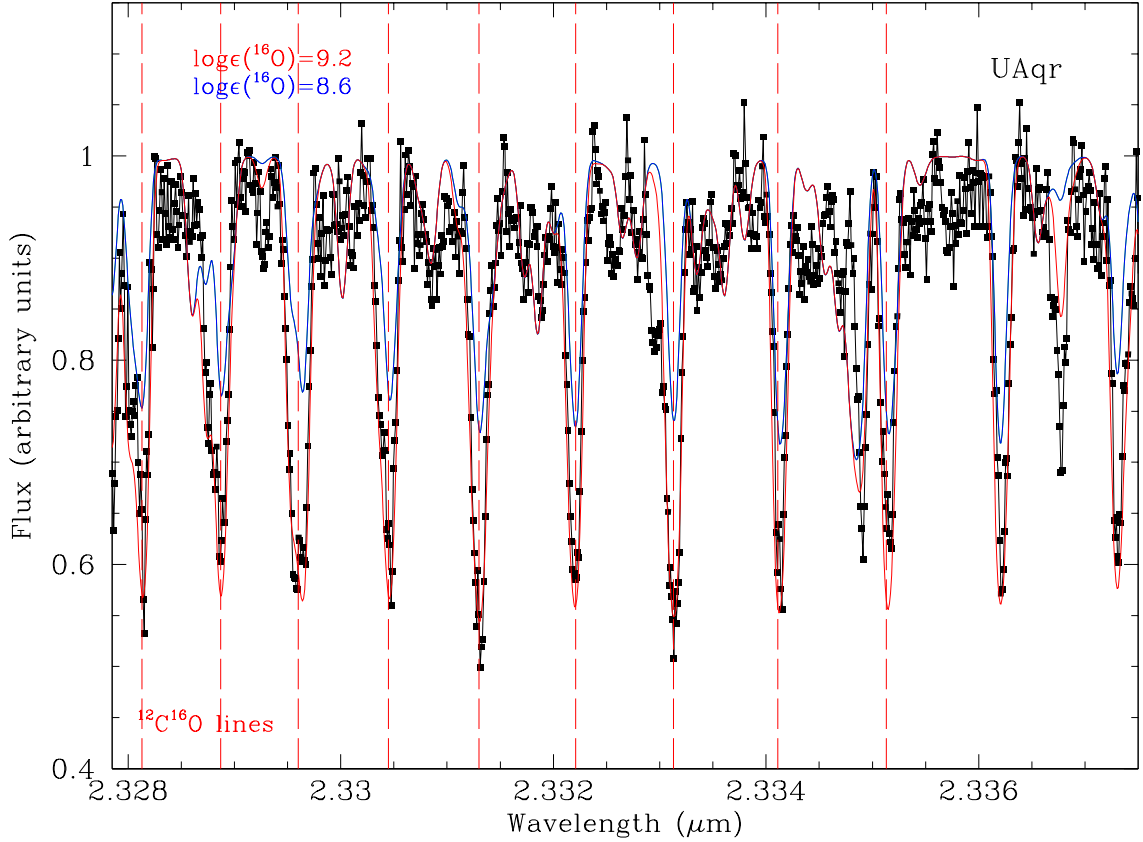}
\caption{Best synthetic (red) and observed (black) spectrum for the wavelength
region centered around 2.333 $\mu$m for the RCB star U Aqr. The red spectrum
assumes T$_{\rm eff}$=6000 K, the input C abundance of 9.52, the N abundance of
9.4, and the $^{16}$O isotopic abundance of log$\varepsilon$($^{16}$O)=9.2. This
region contains $^{12}$C$^{16}$O but not $^{12}$C$^{18}$O lines and the
$^{12}$C$^{16}$O lines are fitted with log$\varepsilon$($^{16}$O)=9.2. The
synthetic spectrum also includes CN lines. A synthetic spectrum (blue) for a
lower $^{16}$O abundance of log$\varepsilon$($^{16}$O)=8.6 is shown for
comparison. Note the strong dependence of the most clear $^{12}$C$^{16}$O lines
(marked with red dashed vertical lines) to changes of the $^{16}$O abundance.
\label{fig6}}
\end{figure}

\clearpage

\begin{figure}
\includegraphics[angle=-90,scale=.60]{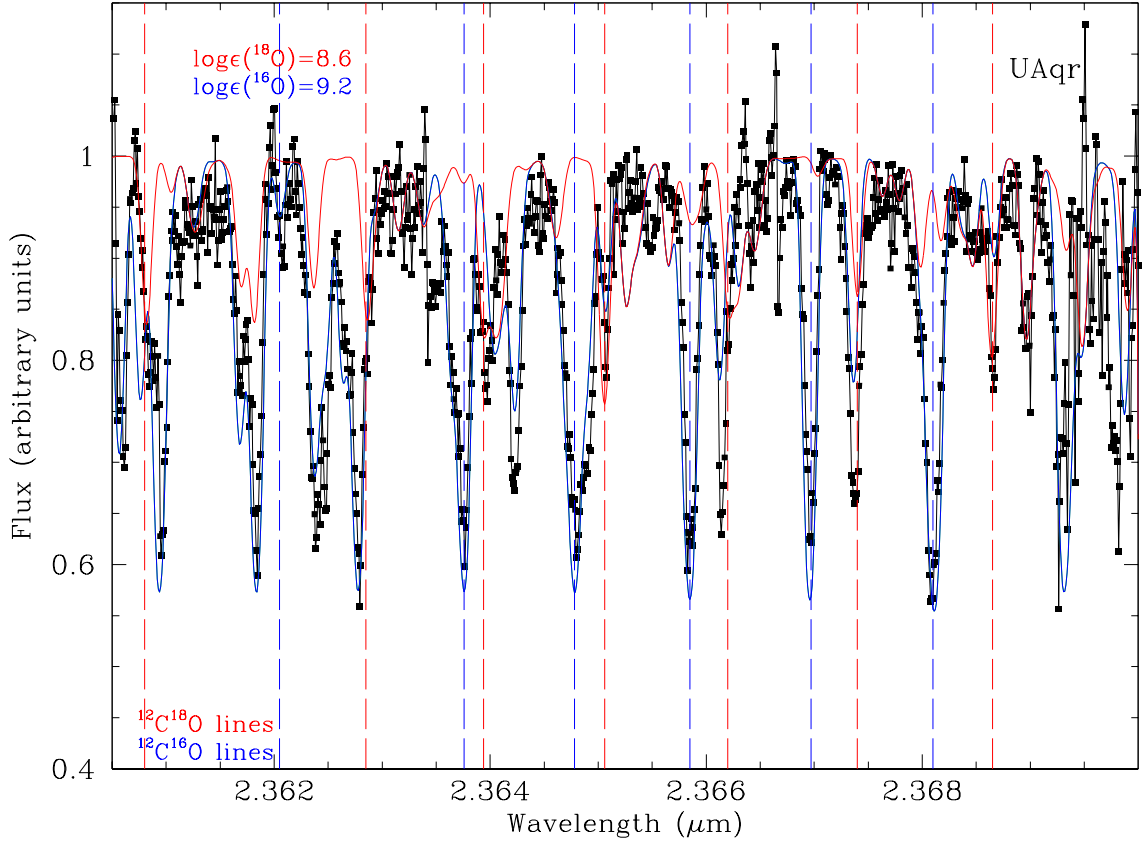}
\caption{Synthetic spectra for the different $^{12}$C$^{16}$O and
$^{12}$C$^{18}$O isomers (blue and red, respectively) and observed (black)
spectrum in the wavelength region centered around 2.365 $\mu$m for the RCB star
U Aqr. Note that the synthetic spectra also include CN lines. The synthetic
spectra are constructed by assuming T$_{\rm eff}$=6000 K, the input C abundance
of 9.52 and using the N abundance of 9.4. The blue and red synthetic spectra
correspond to the oxygen isotopic abundances of log$\varepsilon$($^{16}$O)=9.2
and log$\varepsilon$($^{18}$O)=8.6, respectively (or the isotopic ratio
$^{16}$O/$^{18}$O=4).  
\label{fig7}}
\end{figure}

\clearpage

\begin{figure}
\includegraphics[angle=-90,scale=.60]{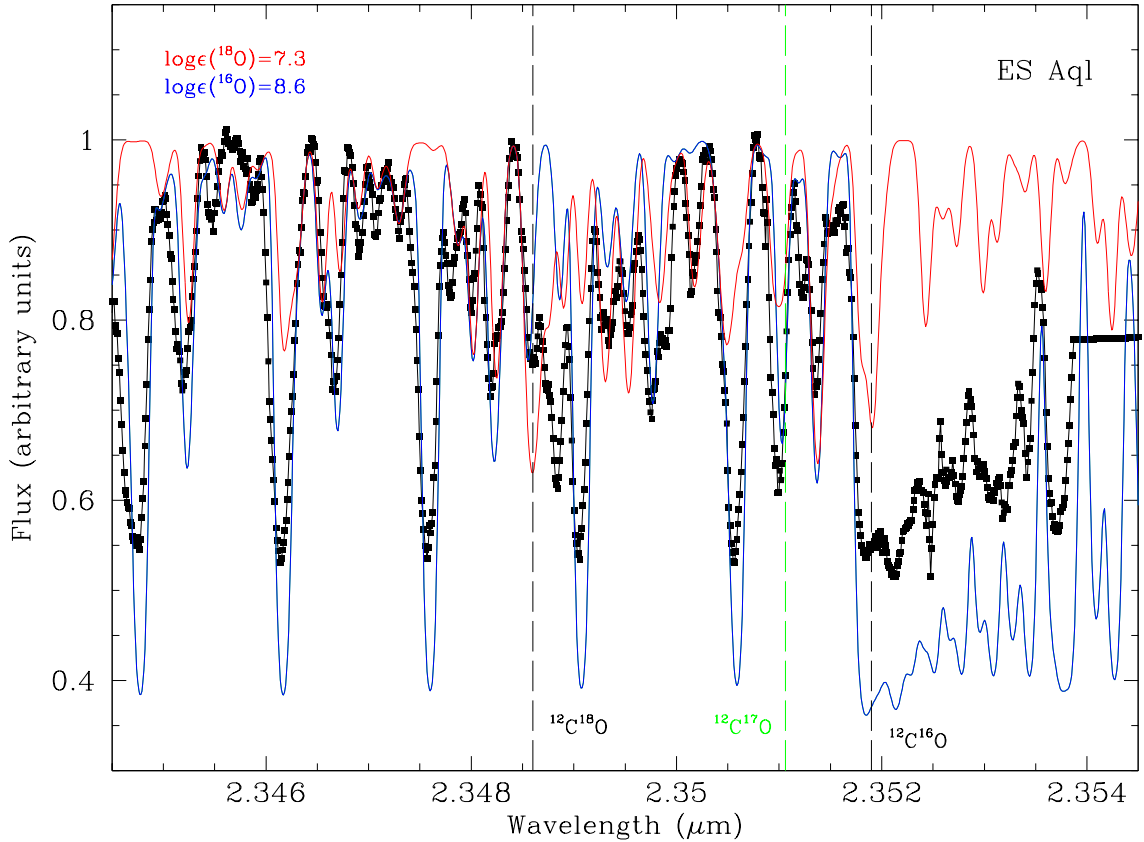}
\caption{Synthetic spectra for the different $^{12}$C$^{16}$O and
$^{12}$C$^{18}$O isomers (blue and red, respectively) and observed (black)
spectrum (on 2008 April 8) in the region around the 2-0 2.349$\mu$m
$^{12}$C$^{18}$O bandhead for the RCB star ES Aql. Note that the synthetic
spectra also include CN lines. The $^{12}$C$^{16}$O lines are saturated for this
star with T$_{eff}$=5000 K. It should be noted that the synthetic blue spectrum
predicts stronger $^{12}$C$^{16}$O lines than observed because the observed
spectrum is diluted by circumstellar dust (see text). The synthetic spectra are
constructed by using the input C abundance of 9.52 and the N abundance of 8.6.
The blue and red synthetic spectra correspond to the oxygen isotopic abundances
of log$\varepsilon$($^{16}$O)=8.6 and log$\varepsilon$($^{18}$O)=7.3,
respectively (or the isotopic ratio $^{16}$O/$^{18}$O=20).
\label{fig8}}
\end{figure}

\clearpage

\begin{figure}
\includegraphics[angle=-90,scale=.60]{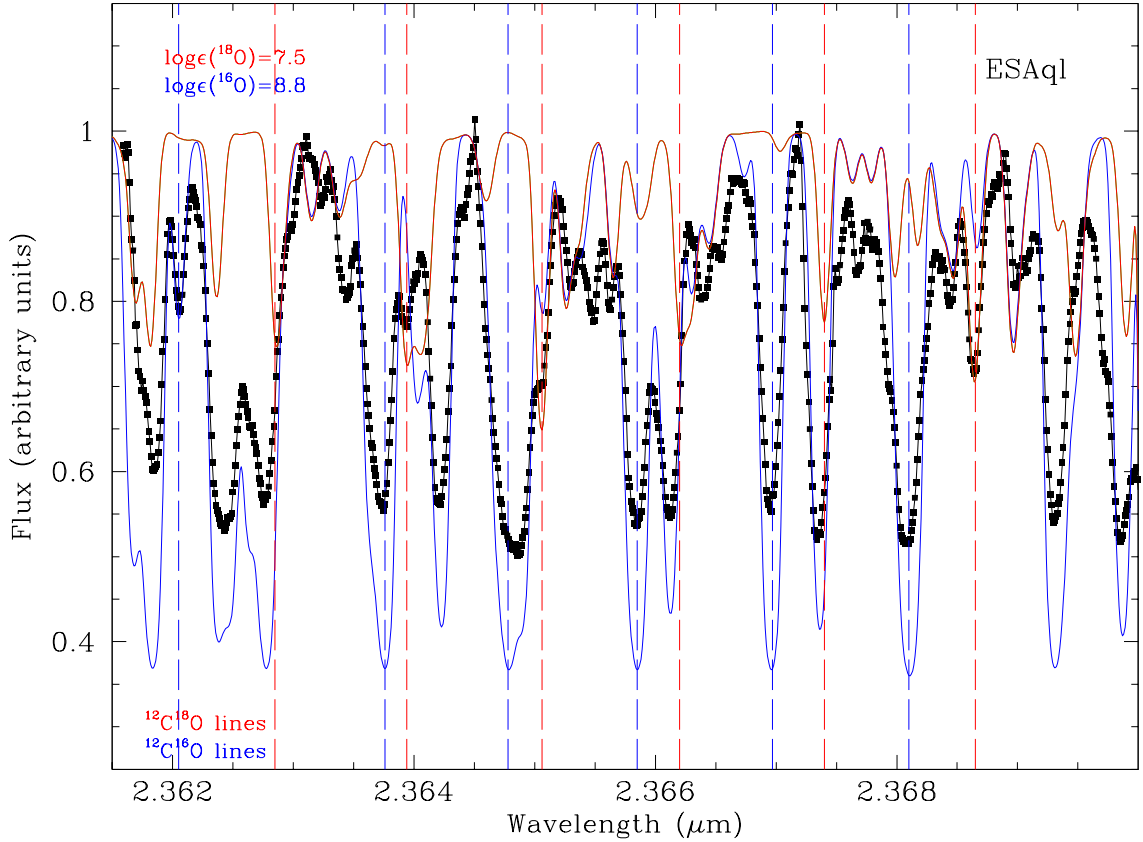}
\caption{Synthetic spectra for the different $^{12}$C$^{16}$O and
$^{12}$C$^{18}$O isomers (blue and red, respectively) and observed (black)
spectrum (on 2008 March 21) in the wavelength region centered around 2.365
$\mu$m for the RCB star ES Aql. Note that the synthetic spectra also include CN
lines. As in Figure 8, the synthetic blue spectrum predicts stronger
$^{12}$C$^{16}$O lines than observed because the observed spectrum is diluted by
circumstellar dust (see text). The synthetic spectra are constructed by using
the input C abundance of 9.52 and the N abundance of 8.6. The blue and red
synthetic spectra correspond to the oxygen isotopic abundances of
log$\varepsilon$($^{16}$O)=8.8 and log$\varepsilon$($^{18}$O)=7.5, respectively
(or the isotopic ratio $^{16}$O/$^{18}$O=20). The relative strength of the weak
$^{12}$C$^{18}$O and $^{12}$C$^{16}$O lines at $\sim$2.3686 $\mu$m and
$\sim$2.3620 $\mu$m provides an estimation of the $^{16}$O/$^{18}$O in this
star. The strongest $^{12}$C$^{18}$O and $^{12}$C$^{16}$O lines - most of them
not useful for abundance estimations - are marked with red and blue vertical
dashed lines, respectively.
\label{fig9}}
\end{figure}

\clearpage

\begin{figure}
\includegraphics[angle=-90,scale=.60]{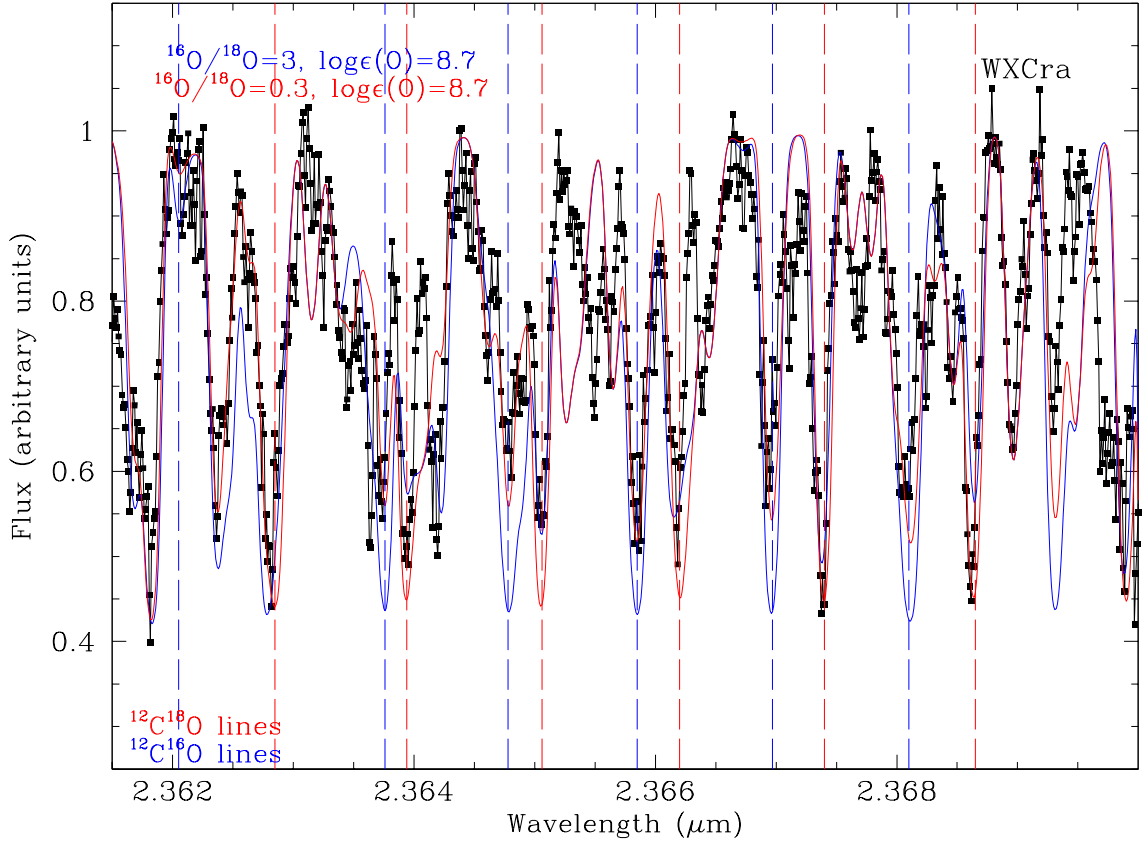}
\caption{Best synthetic (red) and observed (black) spectrum for the region
centered at 2.365 $\mu$m for the RCB star WX CrA. The red spectrum assumes
T$_{\rm eff}$=5300 K, the input C abundance of 9.52, the N abundance of 9.3,  a
total O ($^{16}$O $+$ $^{18}$O) of log$\varepsilon$(O)=8.7 and the isotopic
ratio $^{16}$O/$^{18}$O=0.3. Selected $^{12}$C$^{18}$O and $^{12}$C$^{16}$O lines
are marked with red and blue vertical dashed lines, respectively.  A synthetic
spectrum (blue) for  the same  total O ($^{16}$O $+$ $^{18}$O) of
log$\varepsilon$(O)=8.7  but a higher $^{16}$O/$^{18}$O ratio of 3 (i.e., a
lower $^{18}$O abundance) is shown for comparison. 
\label{fig10}}
\end{figure}

\clearpage

\begin{figure}
\includegraphics[angle=-90,scale=.60]{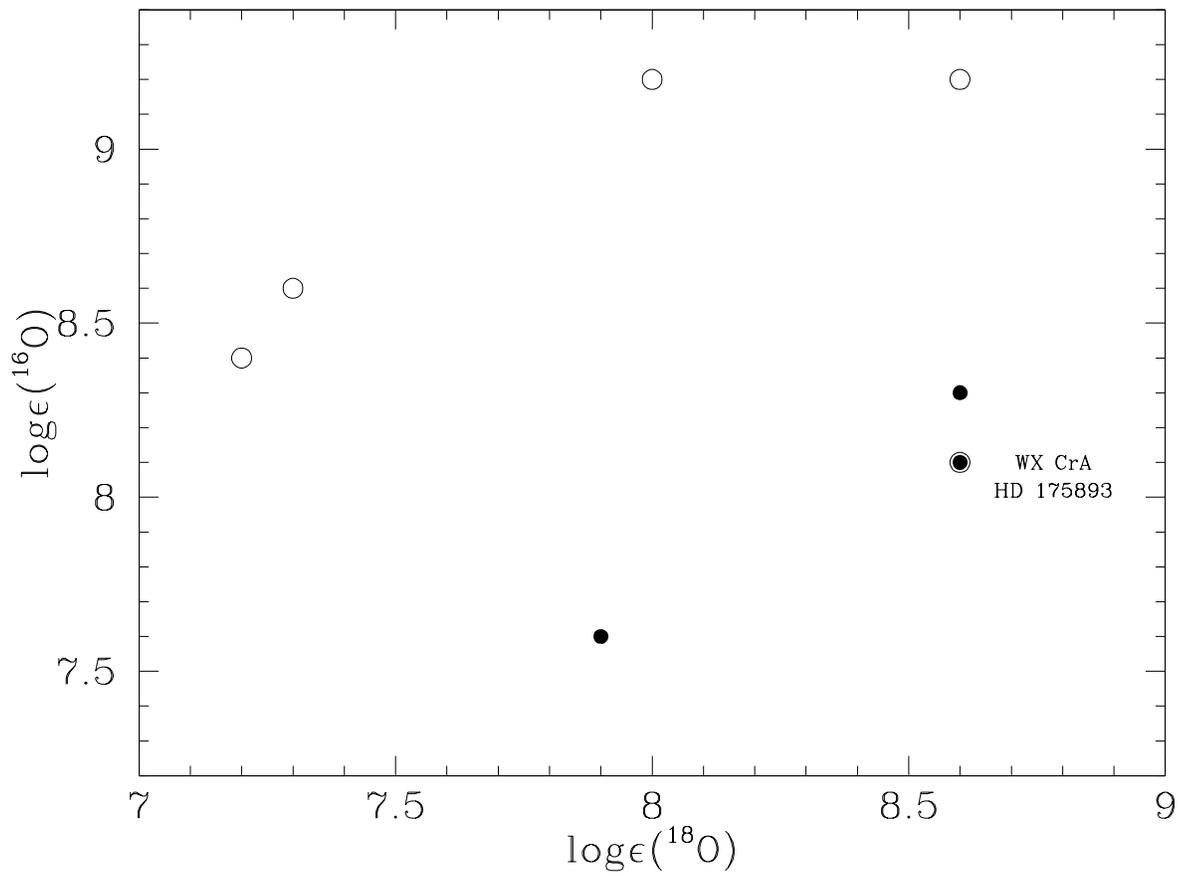}
\caption{The $^{16}$O  abundance versus $^{18}$O abundance for HdC (filled
circles) and cool RCB (unfilled circles) stars. Note that the O isotopic
abundances of the HdC star HD 175893 and the cool RCB WX CrA are identical.
\label{fig11}}
\end{figure}

\end{document}